\documentclass[english]{IEEEtran}
\usepackage[T1]{fontenc}
\usepackage[latin9]{inputenc}
\usepackage{amsthm}
\usepackage{amsmath}
\usepackage{amssymb}
\usepackage{graphicx}

\makeatletter
\numberwithin{equation}{section}
\numberwithin{figure}{section}
\theoremstyle{plain}
\newtheorem{thm}{\protect\theoremname}
\theoremstyle{plain}
\newtheorem{prop}[thm]{\protect\propositionname}
\theoremstyle{plain}
\newtheorem{lem}[thm]{\protect\lemmaname}

\makeatother

\usepackage{babel}
\providecommand{\lemmaname}{Lemma}
\providecommand{\propositionname}{Proposition}
\providecommand{\theoremname}{Theorem}

\begin{document}

\title{Traffic Optimization to Control Epidemic Outbreaks in Metapopulation
Models}

\author{Victor M. Preciado and Michael Zargham}
\maketitle
\begin{abstract}
We propose a novel framework to study viral spreading processes in
metapopulation models. Large subpopulations (i.e., cities) are connected
via metalinks (i.e., roads) according to a metagraph structure (i.e.,
the traffic infrastructure). The problem of containing the propagation
of an epidemic outbreak in a metapopulation model by controlling the
traffic between subpopulations is considered. Controlling the spread
of an epidemic outbreak can be written as a spectral condition involving
the eigenvalues of a matrix that depends on the network structure
and the parameters of the model. Based on this spectral condition,
we propose a convex optimization framework to find cost-optimal approaches
to traffic control in epidemic outbreaks.
\end{abstract}

\section{Introduction}

The development of strategies to control the dynamic of a viral spread
in a population is a central problem in public health and network
security\cite{AM91}. In particular, how to control the traffic between
subpopulations in the case of an epidemic outbreak is of critical
importance. In this paper, we analyze the problem of controlling the
spread of a disease in a population by regulating the traffic between
subpopulations. The dynamic of the spread depends on both the characteristics
of the subpopulation, as well as the structure and parameters of the
transportation infrastructure.

Our work is based on a recently proposed variant of the popular SIS
epidemic model to the case of populations interacting through a network
\cite{MOK09}. We extend this model to a metapopulation framework
in which large subpopulations (i.e., cities) are represented as nodes
in a metagraph whose links represent the transportation infrastructure
connecting them (i.e., roads) \cite{HG97,CV08}. We propose an extension
of the susceptible-infected-susceptible (SIS) viral propagation model
to metapopulations using stochastic blockmodels \cite{HLL83}. The
stochastic blockmodel is a complex network model with well-defined
random communities (or blocks). We model each subpopulation as a random
regular graph and the interaction between subpopulations using random
bipartite graphs connecting adjacent subpopulations. The main advantage
of our approach is that we can find the optimal traffic among subpopulations
to control a viral outbreak solving a standard form convex semidefinite
program.

\section{\label{Notation}Notation \& Preliminaries}

In this section we introduce some graph-theoretical nomenclature and
the dynamic spreading model under consideration.

\subsection{Graph Theory}

Let $\mathcal{G}=\left(\mathcal{V},\mathcal{E}\right)$ denote an
undirected graph with $n$ nodes, $m$ edges, and no self-loops%
\footnote{An undirected graph with no self-loops is also called a \emph{simple}
graph.%
}. We denote by $\mathcal{V}\left(\mathcal{G}\right)=\left\{ v_{1},\dots,v_{n}\right\} $
the set of nodes and by $\mathcal{E}\left(\mathcal{G}\right)\subseteq\mathcal{V}\left(\mathcal{G}\right)\times\mathcal{V}\left(\mathcal{G}\right)$
the set of undirected edges of $\mathcal{G}$. The number of neighbors
of $i$ is called the \emph{degree} of node i, denoted by $d_{i}$.
A graph with all the nodes having the same degree is called regular.
The adjacency matrix of an undirected graph $\mathcal{G}$, denoted
by $A_{\mathcal{G}}=[a_{ij}]$, is an $n\times n$ symmetric matrix
defined entry-wise as $a_{ij}=1$ if nodes $i$ and $j$ are adjacent,
and $a_{ij}=0$ otherwise%
\footnote{For simple graphs, $a_{ii}=0$ for all $i$.%
}. Since $A_{\mathcal{G}}$ is symmetric, all its eigenvalues, denoted
by $\lambda_{1}(A_{\mathcal{G}})\geq\lambda_{2}(A_{\mathcal{G}})\geq\ldots\geq\lambda_{n}(A_{\mathcal{G}})$,
are real. In a regular graph, the largest eigenvalue $\lambda_{1}(A_{\mathcal{G}})$
is equal to the degree of its nodes \cite{Big94}, and the associated
eigenvector is $n^{1/2}\boldsymbol{1}_{n}$ (where $\boldsymbol{1}_{n}$
is the vector of all ones of size $n$).

\subsection{N-Intertwined SIS Epidemic Model}

Our modeling approach is based on the N-intertwined SIS model proposed
by Van Mieghem et at. in \cite{MOK09}. Consider a network of $n$
individuals described by the adjacency matrix $A_{\mathcal{G}}=\left[a_{ij}\right]$.
The infection probability of an individual at node $i\in\mathcal{V\left(G\right)}$
at time $t\geq0$ is denoted by $p_{i}(t)$. Let us assume, for now,
that the viral spreading is characterized by the infection and curing
rates, $\beta_{i}$ and $\delta_{i}$,. Hence, the linearized N-intertwined
SIS model in \cite{MOK09} is described by the following differential
equation: 
\begin{equation}
\frac{d\boldsymbol{p}\left(t\right)}{dt}=\left(BA_{\mathcal{G}}-D\right)\boldsymbol{p}\left(t\right),\label{eq:heteroSIS}
\end{equation}
where $\boldsymbol{p}\left(t\right)=\left(p_{1}\left(t\right),\ldots,p_{n}\left(t\right)\right)^{T}$,
$B=diag(\beta_{i})$, and $D=diag\left(\delta_{i}\right)$. Concerning
the non-homogeneous epidemic model, we have the following result:
\begin{prop}
\label{prop:Heterogeneous SIS stability condition}Consider the heterogeneous
N-intertwined SIS epidemic model in (\ref{eq:heteroSIS}). Then, if
\[
\lambda_{1}\left(BA-D\right)\leq-\varepsilon,
\]
an initial infection $\boldsymbol{p}\left(0\right)\in\left[0,1\right]^{n}$
will die out exponentially fast, i.e., there exists an $\alpha>0$
such that \textup{$\left\Vert p_{i}\left(t\right)\right\Vert \leq\alpha\left\Vert p_{i}\left(0\right)\right\Vert e^{-\varepsilon t}$,
for all $t\geq0$.}
\end{prop}

\section{Spreading Processes in Metapopulations}

\subsection{Metapopulation Model}

Metapopulation models are useful to characterize the dynamics of systems
composed by connected subpopulations \cite{HG97,CV08}. For example,
consider a population of $n$ individuals distributed over $N$ cities
connected via a collection of roads. At a lower level, we can described
the pattern of interactions in the entire population as a massive
graph with $n$ nodes (individuals), where an edge $\{i,j\}$ represents
the interaction between two individuals $i$ and $j$. Alternatively,
we can describe this population at a higher level using a much smaller
graph, called the metagraph, in which nodes represent cities and edges
represent roads connecting cities.

In the metapopulation model, there are two elements to take into consideration.
On the one hand, we have the intrapopulation evolution, which is related
to the evolution of an infection within each subpopulation, as in
isolation. On the other hand, we have the subpopulation interaction,
which is related to encounters between individuals from different
subpopulations. We describe both elements in the following subsections.

\subsubsection{Intrapopulation Connectivity}

Assume we partition the whole population of $n$ individuals into
$N$ subpopulations of sizes $n_{1},...,n_{N}$. We denote the sets
of nodes in each subpopulation by $V_{1},...,V_{N}$.  We also assume
that we are not given any information about the connectivity of individuals
inside each subpopulation, apart from the number of nodes, $n_{i}$,
and the average degree $d_{i}$ of the individuals inside the $i$-th
subpopulation. Hence, it is reasonable to model the connectivity structure
of each subpopulation as a random regular graph of size $n_{i}$ and
degree $d_{i}$. We denote the $n_{i}\times n_{i}$ adjacency matrix
of this random regular graph as $A_{i}$. As we mentioned above, the
largest eigenvalue of this random regular graphs is $\lambda_{1}\left(A_{i}\right)=d_{i}$
and the associated eigenvector is $\boldsymbol{v}_{1}=n_{i}^{1/2}\boldsymbol{1}_{n_{i}}$.

\subsubsection{Subpopulation Interaction}

The interaction between subpopulations is a crucial component that
strongly influences the entire dynamics of the system. To model the
interaction between subpopulations $i$ and $j$, we assume that a
random collection of $w_{ij}$ individuals in $V_{i}$ connect to
a random collection of $w_{ij}$ individuals in $V_{j}$. We can algebraically
represent this connectivity pattern by defining a $n_{i}\times n_{j}$
matrix $A_{ij}$ representing to the structure of a random bipartite
graph connecting two sets of nodes of sizes $n_{i}$ and $n_{j}$
via $w_{ij}$ edges.

\subsubsection{Connectivity matrix of the Population}

The $n\times n$ adjacency matrix of the whole population of individuals
described above is a random matrix, denoted by $\mathcal{A}$, that
can be defined according to block matrices, as follows. First, the
$\left(i,i\right)$-the diagonal block of the population adjacency
matrix is the $n_{i}\times n_{i}$ random matrix $A_{i}$, defined
above. Second, the $\left(i,j\right)$-th off diagonal block is the
$n_{i}\times n_{j}$ matrix $A_{ij}$, defined above. Hence, the adjacency
matrix of the complete population is
\[
\mathcal{A}\triangleq\left[\begin{array}{cccc}
A_{1} & A_{12} & \ldots & A_{1N}\\
A_{21} & A_{2} & \ldots & A_{2N}\\
\vdots & \vdots & \ddots & \vdots\\
A_{N1} & A_{N2} & \ldots & A_{N}
\end{array}\right].
\]

In the following subsection, we apply the N-intertwined SIS epidemic
model to the above adjacency matrix to derive a spectral condition
for stability of a small initial infection.

\subsection{Spreading Dynamics in Metapopulations}

We use (\ref{eq:heteroSIS}) to model the dynamics of an SIS spreading
process in a metapopulations. Assume that the the SIS model spreads
through the individuals in population $V_{i}$ with an spreading rate
$\beta_{i}^{s}$. The recovery rate of individuals in population $V_{i}$
is $\delta_{i}$. We assume that the spreading rate of a virus from
an infected individual in population $j$ towards a susceptible individual
in population $V_{i}$ is equal to $\beta_{i}^{x}$. Let us define
the vector $\boldsymbol{p}_{i}\left(t\right)$ as the $n_{i}$-dimensional
vector containing the probabilities of infection of all the individuals
in the subpopulation $V_{i}$ at time $t\geq0$. Hence, according
to (\ref{eq:heteroSIS}), this vector of infection probabilities evolves
as 
\[
\frac{d\boldsymbol{p}_{i}\left(t\right)}{dt}=\beta_{i}^{s}A_{i}\boldsymbol{p}_{i}\left(t\right)+\sum_{j=1}^{N}\beta_{i}^{x}A_{ij}\boldsymbol{p}_{j}\left(t\right)-\delta_{i}\boldsymbol{p}_{i}\left(t\right),
\]
where the first and last terms represent the spreading and recovery
dynamics within subpopulation $i$. The second term accounts for the
spreading of the disease from subpopulations $j$ to $i$. We can
stack the vectors $\boldsymbol{p}_{i}$ into an $n$-dimensional vector
to write the evolution of the whole population as $\dot{P}\left(t\right)=MP\left(t\right)$,
where $\dot{P}\left(t\right)\triangleq\left[\boldsymbol{p}_{1}^{T}\left(t\right),\ldots,\boldsymbol{p}_{N}^{T}\left(t\right)\right]^{T}$,
and
\begin{eqnarray*}
M & = & \left[\begin{array}{ccc}
\beta_{1}^{s}A_{1}-\delta_{1}I_{n_{1}} & \ldots & \beta_{1}^{x}A_{1N}\\
\beta_{2}^{x}A_{21} & \ldots & \beta_{2}^{x}A_{2N}\\
\vdots & \ddots & \vdots\\
\beta_{N}^{x}A_{N1} & \ldots & \beta_{N}^{s}A_{N}-\delta_{N}I_{n_{N}}
\end{array}\right],
\end{eqnarray*}
According to Proposition \ref{prop:Heterogeneous SIS stability condition},
a small initial infection dies out exponentially fast if the largest
eigenvalue of $M$ is strictly negative. In what follows, we study
the largest eigenvalue of $M$ in terms of metapopulation parameters.
Notice that $M$ is a random matrix, since its blocks represent random
graphs. To analyze the largest eigenvalue of this random matrix, we
make use of the following spectral concentration result from \cite{RCY11}:
\begin{lem}
Consider the random matrix $M$, then, almost surely, 
\[
\left|\lambda_{1}\left(M\right)-\lambda_{1}\left(\mathbb{E}M\right)\right|=o\left(\frac{\log n}{n^{1/2}}\right),
\]
where $\mathbb{E}M$ is the expectation of $M$.

\textup{Using the above lemma, we can find an asymptotic approximation
of $\lambda_{1}\left(M\right)$, for $n\to\infty,$ by computing the
largest eigenvalue of $\mathbb{E}M$. We can compute this eigenvalue
by noticing that $\mathbb{E}A_{i}=\frac{d_{i}}{n_{i}}\boldsymbol{1}_{n_{i}}\boldsymbol{1}_{n_{i}}^{T}$
and $\mathbb{E}A_{ij}=\frac{w_{ij}}{n_{i}n_{j}}\boldsymbol{1}_{n_{i}}\boldsymbol{1}_{n_{j}}^{T}$.
One can then verify that, given the structure of $\mathbb{E}M$, its
largest eigenvalue presents the structure $\boldsymbol{v}_{1}\triangleq\boldsymbol{v}_{1}\left(\mathbb{E}M\right)=\left[\gamma_{1}\boldsymbol{1}_{n_{1}}^{T}\ldots\gamma_{N}\boldsymbol{1}_{n_{N}}^{T}\right]^{T}$.
In particular, the eigenvalue equation is $\mathbb{E}M\ \boldsymbol{v}_{1}=\lambda_{1}\left(\mathbb{E}M\right)\boldsymbol{v}_{1}$,
where}
\[
\mathbb{E}M=\left[\begin{array}{ccc}
\beta_{1}^{s}\frac{d_{1}}{n_{1}}\boldsymbol{1}_{n_{1}}\boldsymbol{1}_{n_{1}}^{T}-\delta_{1}I_{n_{1}} & \ldots & \beta_{1}^{x}\frac{w_{1N}}{n_{1}n_{N}}\boldsymbol{1}_{n_{1}}\boldsymbol{1}_{n_{N}}^{T}\\
\vdots & \ddots & \vdots\\
\beta_{N}^{x}\frac{w_{N1}}{n_{N}n_{1}}\boldsymbol{1}_{n_{N}}\boldsymbol{1}_{n_{1}}^{T} & \ldots & \beta_{N}^{s}\frac{d_{N}}{n_{N}}\boldsymbol{1}_{n_{N}}\boldsymbol{1}_{n_{N}}^{T}-\delta_{N}I_{n_{N}}
\end{array}\right]
\]
and $\lambda_{1}\left(\mathbb{E}M\right)$ is the largest eigenvalue
under study. Or equivalently,
\[
\left(B_{s}\Lambda-\Delta+B_{x}SW\right)\boldsymbol{g}=\lambda_{1}\left(\mathbb{E}M\right)\boldsymbol{g},
\]
where $\boldsymbol{g}=\left[\gamma_{i}\right]$, $B_{s}=diag\left(\beta_{i}^{s}\right)$,
$B_{x}=diag\left(\beta_{i}^{x}\right)$, $\Lambda=diag\left(d_{i}\right)$,
$\Delta=diag\left(\delta_{i}\right)$, $W=W^{T}=\left[w_{ij}\right]$,
and $S=diag\left(1/n_{i}\right)$.
\end{lem}
Therefore, we can approximate the largest eigenvalue of the $n\times n$
matrix $M$ (where $n$ is the number of individuals in the population),
using the largest eigenvalue of the $N\times N$ matrix $B_{s}\Lambda-\Delta+B_{x}SW$
(where $N$ is the number of subpopulations in the model). Hence,
The condition under which the epidemic is guaranteed die out at rate
$\epsilon$ if 
\begin{equation}
\lambda_{\hbox{max}}(B_{s}\Lambda-\Delta+B_{x}SW)\le-\epsilon\label{die}
\end{equation}
is satisfied. In the following section we use this result to find
an optimal distribution of traffic between subpopulations in order
to contain the epidemic spread.

\section{A Convex Framework for Optimal Traffic Control}

\subsection{Traffic Restriction Problem}

We assume that (\ref{die}) is not satisfied without implementing
a travel restriction policy. We define the travel restriction policy
as $\omega=[\omega_{ij}]$ where $0<\omega_{ij}\le W_{ij}$ with cost
$f(\omega)$ convex. The problem travel restriction problem is formally
stated as 
\begin{align}
\min_{\omega}\, & f(\omega)\label{problem}\\
\hbox{\mbox{s.t. }} & \lambda_{\hbox{max}}(B_{s}\Lambda-\Delta+B_{x}S\omega)\le-\epsilon\nonumber \\
 & 0<\omega_{ij}\le W_{ij},\qquad\forall(i,j)\in E\nonumber 
\end{align}
where, since we do not consider permanent relocation between cities,
we assume that $\omega_{ij}=\omega_{ji}$.

The following Lemma states the condition on the model parameters under
which the travel restriction problem is feasible. 
\begin{lem}
\label{feas} There exists a set $\left\{ \omega_{ij}\right\} _{i,j}$
satisfying the constraints in (\ref{problem}) if 
\begin{equation}
\frac{n_{i}}{\beta_{i}^{x}}\left(\epsilon+\beta_{i}^{s}d_{i}-\delta_{i}\right)<0\label{feascond}
\end{equation}
for all cities $i\in{1,\ldots,N}$.\end{lem}
\begin{IEEEproof}
Included in proof of Theorem \ref{sdp}
\end{IEEEproof}
The constraint in Lemma \ref{feas} is equivalent to the condition
that in each individual city the virus would die out at a rate $\epsilon$
with no intercity connections. The virus cannot be forced to die out
in the whole system by controlling intercity connections if it can
persist in any city in isolation.
\begin{thm}
\label{sdp} The traffic restriction problem given in (\ref{problem})
is equivalent to the standard form semidefinite program 
\begin{align}
\min_{\omega}\, & f(\omega)\label{SDP}\\
\hbox{\mbox{s.t. }} & \omega+S^{-1}B_{x}^{-1}(\epsilon I+B_{s}\Lambda-\Delta)\preccurlyeq0\nonumber \\
 & 0<\omega_{ij}\le W_{ij},\qquad\forall(i,j)\in E\nonumber 
\end{align}
\end{thm}
\begin{IEEEproof}
The eigenvalue constraint in (\ref{problem}) is equivalent to 
\begin{equation}
\lambda_{\hbox{max}}(B_{s}\Lambda-\Delta+B_{x}S\omega+\epsilon I)\le0,
\end{equation}
because $I$ can be expressed with any basis. Multiplying by the positive
definite matrix $S^{-1}B_{x}^{-1}$ preserves the sign of the largest
eigenvalue so we can express the relation as 
\begin{equation}
\lambda_{\hbox{max}}\left(S^{-1}B_{x}^{-1}(B_{s}\Lambda-\Delta+\epsilon I)+\omega\right)\le0.\label{sym}
\end{equation}
Since $S^{-1}B_{x}^{-1}(B_{s}\Lambda-\Delta+\epsilon I)+\omega$ is
a symmetric matrix, we can express (\ref{sym}) as the semidefinite
constraint given in (\ref{SDP}), completing the proof of Theorem
\ref{sdp}.

To prove Lemma \ref{feas}, we construct a feasible point $\bar{\omega}$
satisfying the equivalent constraint (\ref{sym}) and the box constraint
$0<\bar{\omega}_{ij}\le W_{ij}$. Let 
\begin{equation}
\xi=\max_{i}\frac{n_{i}}{\beta_{i}^{x}}\left(\epsilon+\beta_{i}^{s}\lambda_{i}-\delta_{i}\right).
\end{equation}
From (\ref{feascond}), $\xi<0$. Applying the triangle inequality
\begin{equation}
\lambda_{\hbox{max}}\left(S^{-1}B_{x}^{-1}(B_{s}\Lambda-\Delta+\epsilon I)+\omega\right)\le\xi+\lambda_{\hbox{max}}(\omega).
\end{equation}
Choosing $\bar{\omega}=\alpha W$ where $0<\alpha\le\min\{1,-\xi/\lambda_{\hbox{max}}(W)\}$
guarantees that $\xi+\lambda_{\hbox{max}}(\bar{\omega})\le0$ and
that $0<\bar{\omega}_{ij}\le W_{ij}$, completing the proof.
\end{IEEEproof}
Standard form semidefinite programs are solvable in polynomial time
via convex optimization methods therefore, a central authority can
set traffic limits on all cities in order to guarantee the epidemic
dies out at rate $\epsilon$ while minimizing the cost.

\section{Local Heuristic Solution}
\begin{thm}
In many cases it may not be possible to compute or implement a centralized
policy. If we suppose the costs are incurred locally by each city
directly effected by the restriction, 
\begin{equation}
f(\omega)=\sum_{i}f_{i}(\omega_{i1},\ldots,\omega_{iN})\label{sep}
\end{equation}
then we can compute a heuristic local solution by first having each
city manager solve 
\begin{align}
U_{i}=\hbox{arg}\min_{u}\, & f_{i}(u)\label{local1}\\
\hbox{s.t.} & \mathbf{1}'u+\frac{n_{i}}{\beta_{i}^{x}}\left(\epsilon+\beta_{i}^{s}\lambda_{i}-\delta_{i}\right)\le0\nonumber \\
 & 0<u_{j}\le W_{ij},\qquad\forall j\nonumber 
\end{align}
then coordinating with neighboring cities allowing traffic 
\begin{equation}
\omega_{ij}=\min\{U_{ij},U_{ji}\}.\label{local2}
\end{equation}
 \label{heur} The Local Heuristic solution defined in (\ref{local1})
and (\ref{local2}) yields a feasible solution to the traffic restriction
problem, (\ref{problem}).\end{thm}
\begin{IEEEproof}
Equation (\ref{local2}) guarantees that $\omega$ is symmetric and
$\omega_{ij}\le U_{ij}$. Combining with the first constraint in (\ref{local1}),
\begin{equation}
\sum_{j}\omega_{ij}+\frac{n_{i}}{\beta_{i}^{x}}\left(\epsilon+\beta_{i}^{s}\lambda_{i}-\delta_{i}\right)\le0.\label{ddom}
\end{equation}
The box constraint in (\ref{local1}) guarantees that $0<\omega_{ij}\le W_{ij}$.
Consider the matrix 
\begin{equation}
\omega+S^{-1}B_{x}^{-1}(\epsilon I+B_{s}\Lambda-\Delta),\label{mat}
\end{equation}
whose diagonal entries are strictly negative according to Lemma \ref{feas}.
Equation (\ref{ddom}) guarantees that the matrix (\ref{mat}) is
diagonally dominant. Theorem 6.1.10 from \cite{HJ85} guarantees that
the matrix in (\ref{mat}) is negative semidefinite, satisfying the
eigenvalue constraint and completing the proof.
\end{IEEEproof}
The heuristic solution proposed in (\ref{local1}) and (\ref{local2})
is local in the sense that traffic restrictions on the edges in the
intercity network can be computed by each city solving the optimal
restrictions for the edges connecting them to other cities. Two cities
will not necessarily compute the same optimal restriction so the minimum
of the two values is used. This is consistent with our model because
all travel assumed to be is round trip, thus the realized traffic
can be at most the minimum of the traffic allowed by the two cities
involved. Theorem \ref{heur} formally guarantees that this local
method yields a solution that causes the virus to die out at at least
rate $\epsilon$.

\section{Numerical Experiments}

\begin{figure}
\includegraphics[width=0.75\columnwidth]{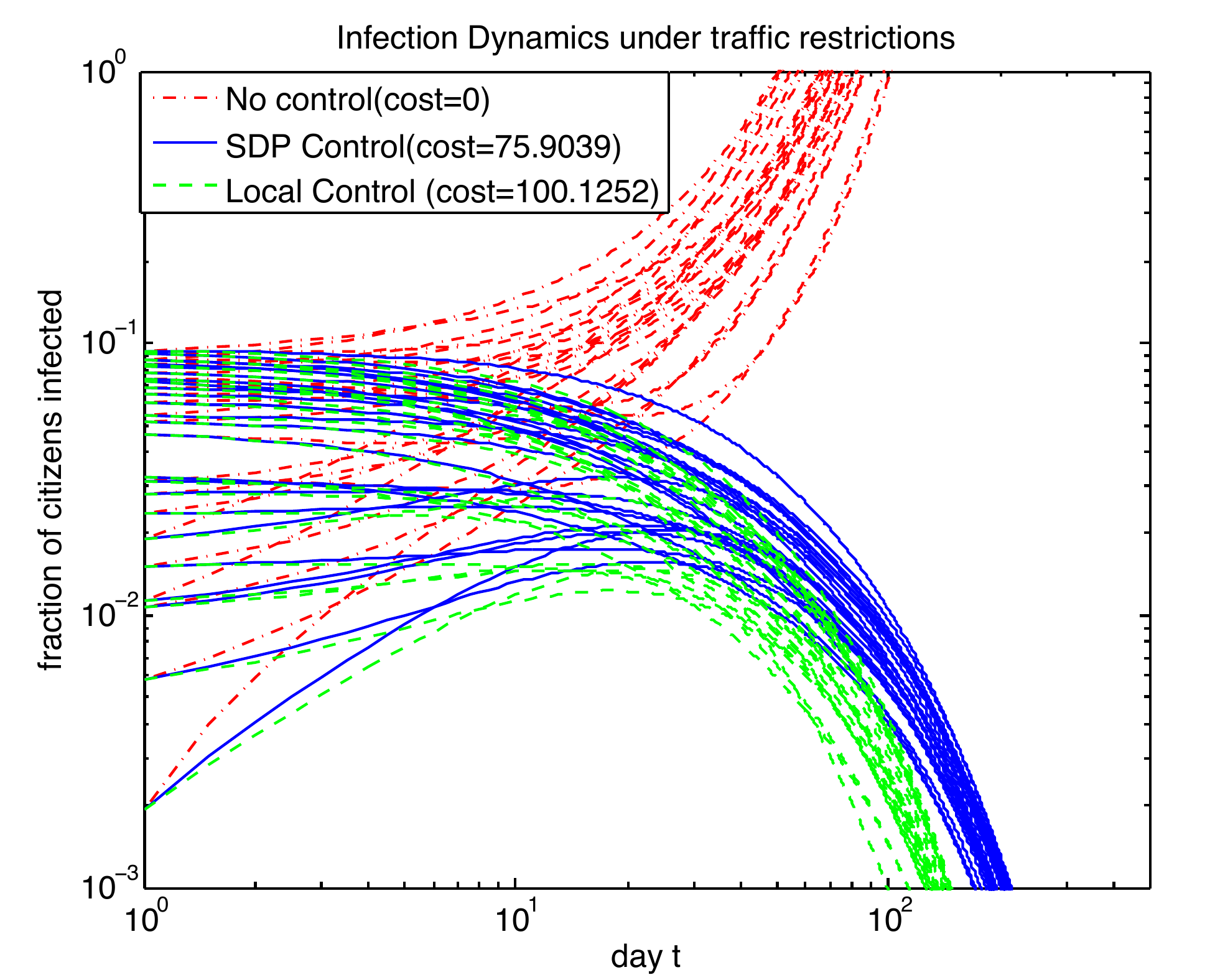} \centering
\caption{\label{sim}The fraction of people who are infected is driven to 0
in every city via traffic control. The cost incurred by the local
heuristic is about 33\% greater than the optimal cost. }
\end{figure}

We demonstrate the relative performance of our centralized and local
solutions using a sample problem with randomly generated parameters.
We choose the cost function 
\begin{equation}
f(\omega)=-\sum_{(i,j)\in E}\ln\left(\frac{\omega_{ij}}{W_{ij}}\right)\label{cost}
\end{equation}
because it is convex and satisfies (\ref{sep}). Furthermore, (\ref{cost})
intuitively captures the cost of restricting traffic on each edge
because there is no cost when traffic is unrestricted (i.e. $w_{ij}=W_{ij}$)
but the cost tends to $\infty$ as $w_{ij}\to0$. It is not possible
to completely delete an intercity connection.

Figure \ref{sim} shows that with no traffic restrictions all cities
go to 100\% infection rate while both the optimal solution and local
heuristic force the virus to die out. The heuristic solution incurs
a higher total cost but also forces the epidemic to die out faster.

\begin{figure}
\includegraphics[width=1\columnwidth]{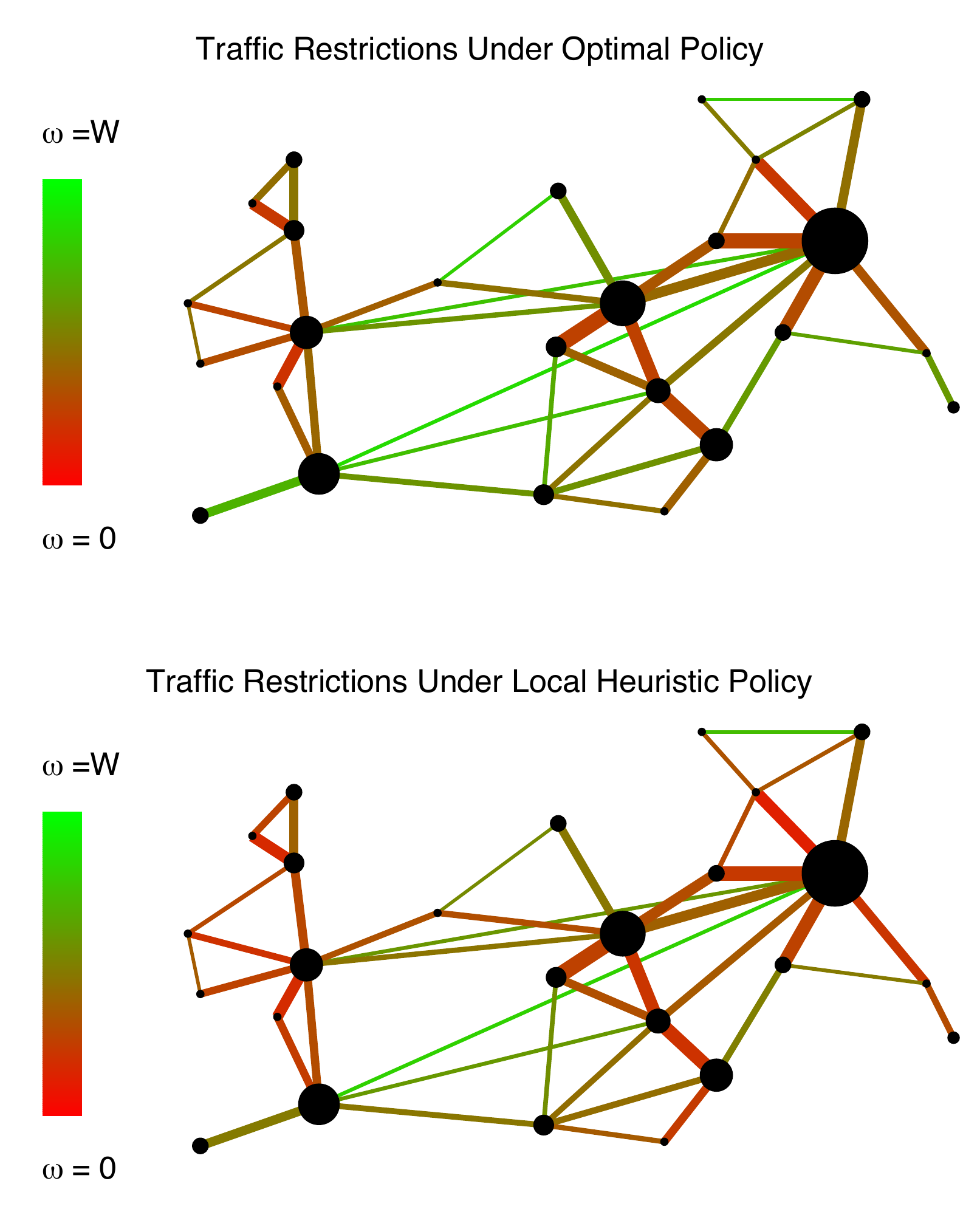} \centering
\caption{\label{net} More traffic restrictions are required by the heuristic
solution, however there is a strong correlation between which edges
are restricted by the heuristic solution and by the optimal solution. }
\end{figure}

Figure \ref{net} shows the network of cities and the traffic restrictions
on the edges. Points representing cities are scaled proportional to
their populations $n_{i}$. Edges are scaled proportional to the the
unrestricted traffic $W_{ij}$. The color of each edge is linearly
scaled from green ($\omega_{ij}=W_{ij}$) to red ($\omega_{ij}=0$).
It significant to note that despite the local approach, the restrictions
imposed are very similar to the optimal case.

\section{Conclusions}

We have proposed a convex framework to contain the propagation of
an epidemic outbreak in a metapopulation model by controlling the
traffic between subpopulations. In this context, controlling the spread
of an epidemic outbreak can be written as a spectral condition involving
the eigenvalues of a matrix that depends on the network structure
and the parameters of the model. Based on our spectral condition,
we can find cost-optimal approaches to traffic control in epidemic
outbreaks by solving an efficient semidefinite program.

\end{document}